\documentclass[11pt]{article} 

\usepackage{fullpage, graphicx, geometry, microtype, url, colortbl, titlesec, ragged2e, float}
\usepackage{amsmath, amsthm, amsfonts, amssymb, mathtools, bm, longtable, adjustbox, enumitem, setspace}
\usepackage[skip = 10pt, indent = 20pt]{parskip}
\usepackage[hidelinks]{hyperref} 
\usepackage{hyperref, booktabs, siunitx, array, caption, subcaption, multirow}
\usepackage{algorithm, algpseudocode}
\usepackage[dvipsnames]{xcolor}
\usepackage[all]{hypcap}

\hypersetup{
    colorlinks = true,   
    linkcolor = blue,    
    urlcolor = blue,     
    citecolor = red      
}
\usepackage{tikz}
\usepackage[object=vectorian]{pgfornament}
\usepackage{xcolor} 
\newcommand{\fancyline}{
  \begin{center}
    \begin{tikzpicture}[scale=0.5]
      \node at (0,0) {\pgfornament[width = 8cm, color = gray]{88}};
    \end{tikzpicture}
  \end{center}
}

\definecolor{sectioncolor}{HTML}{1F3B73}    
\definecolor{subsectioncolor}{HTML}{006400}    
\definecolor{subsubsectioncolor}{HTML}{8B4513} 

\titleformat{\section}
  {\color{sectioncolor}\normalfont\Large\bfseries}  
  {\thesection}{1em}{}
\titleformat{\subsection}
  {\color{subsectioncolor}\normalfont\large\bfseries} 
  {\thesubsection}{1em}{}
\titleformat{\subsubsection}
  {\color{subsubsectioncolor}\normalfont\large\bfseries} 
  {\thesubsubsection}{1em}{}

\renewcommand{\arraystretch}{1.2}

\newgeometry{
    top = 0.9 in,
    bottom = 0.9 in,
    outer = 0.9 in,
    inner = 0.9 in
}

\title{\textbf{\LARGE Novel Risk Measures for Portfolio Optimization \\ Using Equal-Correlation Portfolio Strategy}}

\author{
    Biswarup Chakraborty\thanks{I thank Prof. Diganta Mukherjee for his guidance and advice.} \\
    \normalsize Indian Statistical Institute, Kolkata \\
    \normalsize \hypersetup{urlcolor=black}\href{mailto:biswarupchakraborty891@gmail.com}{\texttt{biswarupchakraborty891@gmail.com}}
}

\date{} 

\begin{document}
\maketitle

\renewcommand{\abstractname}{\large Abstract}
\begin{abstract}

Portfolio optimization has long been dominated by covariance-based strategies, such as the Markowitz Mean-Variance framework. However, these approaches often fail to ensure a balanced risk structure across assets, leading to concentration in a few securities.

In this paper, we introduce novel risk measures grounded in the \textbf{equal-correlation portfolio strategy}, aiming to construct portfolios where each asset maintains an equal correlation with the overall portfolio return. We formulate a mathematical optimization framework that explicitly controls portfolio-wide correlation while preserving desirable risk-return trade-offs.

The proposed models are empirically validated using historical stock market data. Our findings show that portfolios constructed via this approach demonstrate superior risk diversification and more stable returns under diverse market conditions.

This methodology offers a compelling alternative to conventional diversification techniques and holds practical relevance for institutional investors, asset managers, and quantitative trading strategies.
\end{abstract}
\newpage
\setcounter{footnote}{0}
\section{Introduction} \label{sec:introduction}

\hspace{20pt}Portfolio optimization is a cornerstone of financial decision-making, aiming to construct portfolios that balance risk and return effectively~\cite{demiguel2009optimal}. Traditional models, such as the Markowitz Mean-Variance framework, rely on covariance-based risk assessment, often leading to portfolios heavily influenced by a few dominant assets. However, these models fail to account for correlation structures~\cite{narayan2023correlation} explicitly, which can result in suboptimal diversification.

\begin{figure}[H]
    \centering
    \includegraphics[width = 0.93\linewidth]{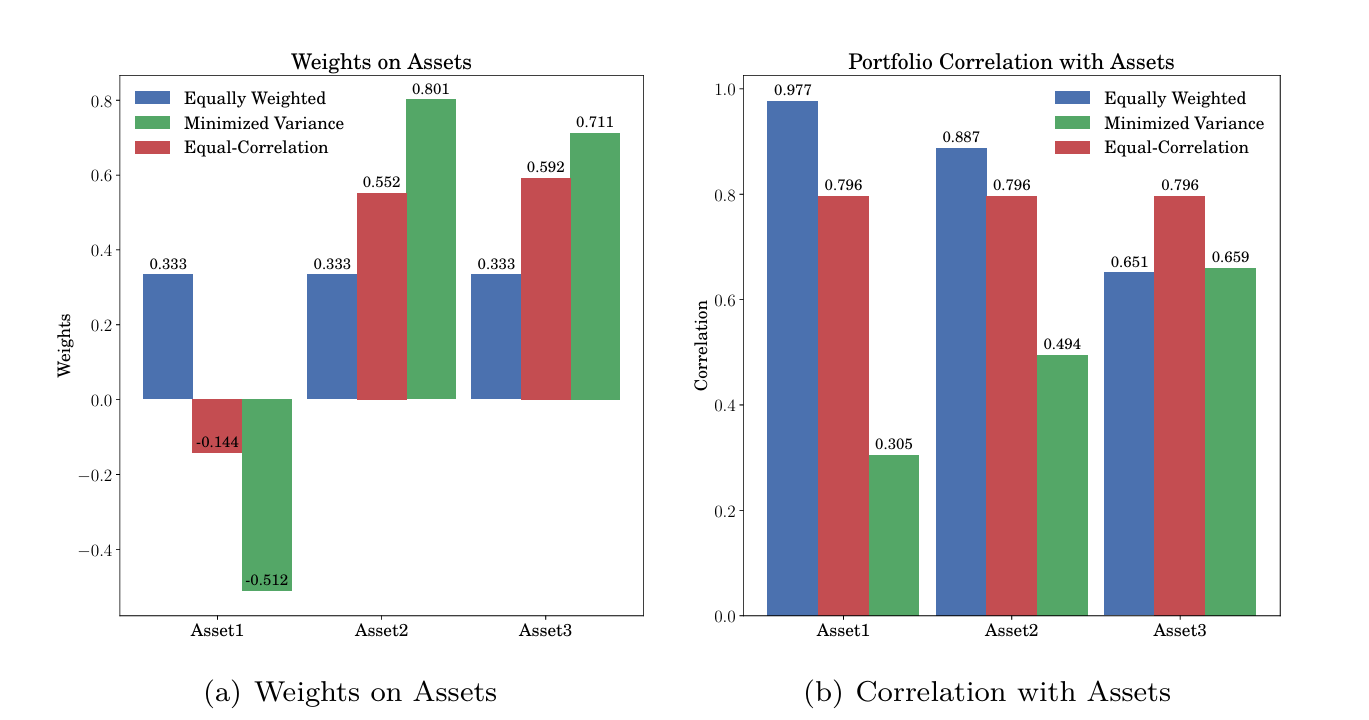}
    \caption{Comparison of Three Strategies}
    \label{fig:intro_weight_plot}
\end{figure}

An alternative approach, \textbf{equal-correlation portfolio strategies}, seeks to allocate weights such that all assets in the portfolio maintain an equal correlation with the overall portfolio. This method offers better risk dispersion and enhances stability, especially in volatile market conditions. Figure \ref{fig:intro_weight_plot} which was obtained from paper~\cite{Jia2024} gives the comparison of weights and correlations among three standard portfolio strategies: \textbf{equally-weighted}, \textbf{equal-correlation}, and \textbf{minimum variance portfolios}. The figure shows that the minimum variance portfolio is aggressive with a negative weight on the first asset and implies a concentration in terms of weights on the latter two assets.

On the other hand, due to the covariance between assets, the equally weighted strategy does not perform well enough~\cite{demiguel2009optimal} in terms of correlation with component assets, although they average out in terms of weights. Compared to the other two strategies, the equal-correlation strategy gets an absolute average in terms of correlation with component assets and is a compromise between the other two in terms of weights on assets. In this paper, instead of working with the equal-correlation portfolio strategies, we will define some new measures based on the equal-correlation strategy. Our analysis is structured into 4 key segments.

In section~\ref{sec:equal-correlation-measures} We introduce \textbf{new risk measures} based on the equal-correlation portfolio strategy and derive their formulas. Section~\ref{sec:portfolio-optimization-models} introduces the \textbf{portfolio optimization models} used for bench-marking, including both conventional approaches and newly proposed strategies. Section~\ref{sec:methodology} outlines the \textbf{methodology} for \textbf{simulating} and evaluating \textbf{portfolio performance} over the period 2010–2019, detailing the data processing steps and model implementation. Finally, in section~\ref{sec:portfolio-optimization-results} We take a look at our \textbf{stock universe}, present the \textbf{empirical results}, offer a interpretations of key findings, and discuss their implications for portfolio management. Then we look at the \textbf{conclusions} (section~\ref{sec:conclusion})

All simulations were conducted using \textbf{Python}, leveraging robust numerical and optimization libraries. Daily stock price data was sourced via the \textbf{yfinance}~\cite{yfinance} package, ensuring a comprehensive dataset for backtesting and performance evaluation. Our dataset comprises daily stock returns of about \textbf{350} stocks which are part of \textbf{S\&P500} from 2009 to 2019. Here, we will be working with \textbf{Gross \% Returns}\footnote{Gross Returns of a stock for today is calculated as (Throughout this paper, unless mentioned otherwise, stock return means gross \% return of the stock)\[\text{Gross \% Return} = 100\times\frac{\text{Today's Closing Price}}{\text{Previous Day's Closing Price}}\]} of the stocks.

\section[Risk Measures Based on Equal-Correlation Strategy]{Risk Measures Based on Equal-Correlation Strategy} \label{sec:equal-correlation-measures}

\hspace{20pt}Before introducing the risk measures, we will first examine the Equal-Correlation Portfolio Strategy, which serves as the cornerstone of our discussion.

\subsection{Equal-correlation Weight Vector (\texorpdfstring{$\bm{w_{eq}}$}{weq})} \label{subsec:equal-correlation-weight-vector}

\hspace{20pt}Let's take a look at \textbf{Equal-correlation weight vector}~\cite{Jia2024}. Let us consider $\bm{d}$ assets\footnote{\normalsize Throughout this paper, we will use the terms assets and stocks interchangeably} with a returns vector $\tilde{r} \in \mathbb{R}^d$ representing their \textbf{gross \% returns}, where $r_i$ denotes the return of the $i$-th stock and their symmetric, positive definite covariance matrix $\Sigma \in \mathbb{R}^{d \times d}$. We use $w \in \mathbb{R}^d$ to denote the \textbf{portfolio weight vector}, this represents the proportion of total capital allocated across different stocks.  The weights are readily interpreted. $w_i = 0.05$ means that 5\% of the total portfolio value is held in asset $i$, and $w_k = -0.01$ means that we hold a short position in asset $k$, with value 1\% of the total portfolio value. So, $\sum_{i=1}^{d} w_i = 1$, that is, $\mathbf{1}^T w = 1$ ($\mathbf{1}$ denotes a vector where all elements are equal to one). The correlation between the portfolio return $w^T \tilde{r}$ and the return of the $i$-th asset $r_i$ is expressed by ($Std$ denotes the standard deviation)
\begin{equation} \label{eq:corr-vec}
    \text{Corr}(w^T \tilde{r}, r_i) = \frac{\text{Cov}(w^T \tilde{r}, r_i)}{\text{Std}(w^T \tilde{r}) \text{ Std}(r_i)} = \frac{w^T \Sigma e_i}{\sqrt{w^T \Sigma w} \text{ Std}(r_i)}
\end{equation}
Where $e_i = (00 \dots 010 \dots 00)^T \in \mathbb{R}^d$ (the 1 being in the $i$ -th position). Through this, we can obtain the weight vector of the equal correlation portfolio vector $w_{eq}$ that satisfies the criteria,
$$\frac{w_{eq}^{T} \Sigma e_i}{\text{ Std}(r_i) \sqrt{w_{eq}^{T} \Sigma w_{eq}}} = \frac{w_{eq}^{T} \Sigma e_j}{\text{ Std}(r_j) \sqrt{w_{eq}^{T} \Sigma w_{eq}}}, \quad \forall i, j \in \{1,2,\dots,d\}, i\neq j$$
This is equivalent to $\Sigma w_{eq} \propto \text{Std}(\tilde{r})$, where $\text{Std}(\tilde{r})$ is the column vector of the standard deviation of the returns of the stocks.
$$\Sigma w_{eq} = k \times \text{Std}(\tilde{r}) \ [k > 0] \Rightarrow \quad w_{eq} = \ k \Sigma^{-1} \text{Std}(\tilde{r})$$
$$\because \mathbf{1}^T w_{eq} = 1, \quad k \times \mathbf{1}^T \Sigma^{-1} \text{Std}(\tilde{r}) = 1 \Rightarrow \quad k = \frac{1}{\mathbf{1}^T \Sigma^{-1} \text{Std}(\tilde{r})}$$
\begin{equation} \label{eq:w-eq}
    \Rightarrow w_{eq} = \frac{\Sigma^{-1} \text{Std}(\tilde{r})}{\mathbf{1}^T \Sigma^{-1} \text{Std}(\tilde{r})}
\end{equation}

However, since this gives a \textbf{unique} weight vector, we can not combine this with other measures of portfolio quality (e.g, mean return, portfolio variance). To improve this, we define new measures, which we can use to assess the extent to which the equal-correlation condition is being satisfied. We define some \textbf{new measures} below:

\subsection{Square Distance from \texorpdfstring{$\bm{w_{eq}}$}{weq} (\texorpdfstring{$\bm{d_{eq}^2}$}{d2weq})} \label{subsec:square-distance}

\hspace{20pt}Here, we define the squared Euclidean distance from $\bm{w_{eq}}$ as a stability measure, a concept supported by Kim et al.~\cite{kim2022distance}, who demonstrate its effectiveness in portfolio evaluation by measuring similarity to a reference portfolio. Our reasoning is that a portfolio is more desirable when its weight vector is closer to the equal-correlation portfolio vector. Hence, a lower value indicates a more stable portfolio. Specifically, for any given portfolio weight vector $w$, we define the risk measure as $\bm{d_{eq}^2=||w - w_{eq}||_2^2}$, where $||.||_2$ represents the $L_2$-norm\footnote{\normalsize For a vector $\tilde{x}$, $$||\tilde{x}||_2 = \sqrt{\tilde{x}^T\tilde{x}}$$} of a vector.

\subsection{Correlation Variance \texorpdfstring{($\bm{\sigma_{\rho}^2}$)}{srho}} \label{subsec:correlation-variance}

\hspace{20pt}This measure is more refined as it explicitly captures the \textbf{variation in the correlation} between \textbf{portfolio returns} and the returns of its \textbf{constituent stocks}. Consider a set of $d$ assets with a return vector $\tilde{r} \in \mathbb{R}^d$ representing their returns, where $r_i$ denotes the return of the $i$-th stock. The portfolio weight vector is given by $w \in \mathbb{R}^d$. Consequently, the portfolio returns are expressed as $w^T \tilde{r}$. We analyze the correlations between the portfolio returns and the returns of its component stocks, forming a $d$-dimensional vector. Here, $\Sigma$ represents the covariance matrix of stock returns, and $Std$ denotes the standard deviation.
$$\bm{corr\_vec} = \frac{w^T \Sigma e_i}{\sqrt{w^T \Sigma w}\text{ Std}(r_i)} \quad \text{(from equation \ref{eq:corr-vec})}$$
Then we consider $\bm{\sigma_{\rho}^2}$ as the variance of $\bm{corr\_vec}$ as a measure of \textbf{risk} associated with the portfolio, since higher values indicate significant variation in the correlations between the portfolio returns and the returns of its constituent stocks.

\subsubsection{Derivation of the formula for \texorpdfstring{$\bm{\sigma_{\rho}^2}$}{srho}} \label{subsubsec:derivation-of-formula}

\hspace{20pt}Let $\bm{N}$ denote the number of training days, $\bm{d}$ the number of constituent stocks in our portfolio. So, our \textbf{training period} (subsection \ref{subsec:training-testing-period}) daily stock returns can be represented as a matrix $\bm{X_{N\times d}}$, where the $(i,j)$-th element denotes return of $j$-th stock on $i$-th training day. $w \in \mathbb{R}^d$ be the portfolio weight vector ($1^T w = 1$). So, portfolio returns vector is $\bm{X w} \in \mathbb{R}^N$.($\bm{1_N}$ denotes the $N$-dimensional vector of ones, $\bm{J_N}$ denotes the $N$-dimensional matrix with all its elements equal to 1 and $\bm{I_N}$ denotes the identity matrix of order $N\times N$)
\vspace{0.3cm}
\\Mean value of daily returns is $\frac{1}{N}X^T 1_N$
\\Centered Daily return is $X - \frac{1}{N}1_N 1_N^T X = (I_N - \frac{1}{N} J_N)X$
\\Similarly, centered Portfolio Returns is $(I_N - \frac{1}{N} J_N)X w$
\\Covariance matrix of the daily returns = $\frac{1}{N}[(I_N-\frac{1}{N}J_N)X]^T [(I_N-\frac{1}{N}J_N)X]$
$$
= \frac{1}{N}X^T(I_N-\frac{1}{N}J_N)(I_N-\frac{1}{N}J_N)X
$$
$$
= \frac{1}{N}X^T(I_N-\frac{2}{N}J_N+\frac{1}{N}J)\quad [J_N^2=NJ_N]
$$
$$
= \frac{1}{N}X^T(I_N-\frac{1}{N}J_N)X
$$
We will replace this with some \textbf{positive definite shrinkage estimate}~\cite{boas2017shrinkage} say $S\in\mathbb{R}^{d\times d}$.

If the sample covariance matrix is $\hat{\Sigma}$, and F is some well structured estimate of $\Sigma$, in our case, we take F as the constant correlation covariance matrix, which is an estimate of the covariance matrix, assuming that correlation between any two stocks is equal, the the shrinkage estimate is defined as 
$$
S = (1-\lambda) \hat{\Sigma} + \lambda F,\quad (\lambda \in (0,1))
$$
\hspace{20pt}Different methods discussed in subsection \ref{subsec:covariance-matrix-mean-estimate} provide different estimates of $\lambda$. Similarly, we get portfolio returns variance as $\frac{1}{N}[(I_N-\frac{1}{N}J_N)Xw]^T [(I_N-\frac{1}{N}J_N)Xw]$
$$
= \frac{1}{N}w^TX^T(I_N-\frac{1}{N}J_N)Xw
$$
\hspace{20pt}Instead of using the above form, we'll use $w^TSw$, since $S$ is some shrinkage estimate for $\frac{1}{N}X^T(I_N-\frac{1}{N}J_N)X$. And the covariance between daily returns and portfolio returns is $\frac{1}{N}[(I_N-\frac{1}{N}J_N)X]^T [(I_N-\frac{1}{N}J_N)Xw]$
$$
= \frac{1}{N}X^T(I_N-\frac{1}{N}J_N)Xw
$$
\hspace{20pt}Here, we also use the vector $Sw \in \mathbb{R}^{d}$. Since $S$ serves as the estimate for the covariance matrix, the standard deviations of stock returns are given by the square root of its diagonal elements. Let $\bm{\Lambda}$ be the diagonal matrix where the diagonal elements match those of $S$, while all off-diagonal elements are zero. Then, the matrix containing the standard deviations of stock returns along the diagonal is denoted as $\bm{\Lambda^{\frac{1}{2}}}$ (\textbf{element-wise square root} of the $\bm{\Lambda}$ matrix). Thus, the vector representing the correlation terms between stock returns and the portfolio return is given by:
$$
\Lambda^{-\frac{1}{2}} Sw \frac{1}{\sqrt{w^TSw}} \in \mathbb{R}^{d}
$$
Now if we consider the variance of this vector, we get the form,
$$
\sigma_{\rho}^2 = \left(\frac{1}{d}\right)\frac{(\Lambda^{-\frac{1}{2}} Sw)^T (I_d - \frac{1}{d} J_d)(\Lambda^{-\frac{1}{2}} Sw)}{w^TS w}
$$
$$
= \left(\frac{1}{d}\right)\frac{w^TS\Lambda^{-\frac{1}{2}} (I_d - \frac{1}{d} J_d)\Lambda^{-\frac{1}{2}} Sw}{w^TS w}
$$

Here, we have used the fact that $\Lambda$ and $S$ both are symmetric matrices. $S$ and $\Lambda$ both are positive definite matrices and $(I_d - \frac{1}{d} J_d)$ is an idempotent matrix\footnote{Matrix \textbf{A} is idempotent implies $\bm{A^2=A}$ and so all \textbf{eigen values} are either 0 or 1} and hence it's non-negative definite. So, the matrix $S\Lambda^{-\frac{1}{2}}(I_d - \frac{1}{d} J_d)\Lambda^{-\frac{1}{2}}S$ is a non-negative definite matrix. So, for any $w\in \mathbb{R}^{d}$, the numerator of $\sigma_{\rho}^2$ takes non-negative value, and the denominator takes positive value. So, the value of $\sigma_{\rho}^2$ exists for any $w \in \mathbb{R}^d$. For future reference, we'll denote,
\begin{equation} \label{eq:corr-var}
    \sigma_{\rho}^2(w) = \left(\frac{1}{d}\right)\frac{w^TS\Lambda^{-\frac{1}{2}} (I_d - \frac{1}{d} J_d)\Lambda^{-\frac{1}{2}} Sw}{w^TS w}
\end{equation}

\section{Portfolio Optimization Models} \label{sec:portfolio-optimization-models}

\hspace{10pt}Here, we define the portfolio optimization models. We will also be using \textbf{Markowitz portfolio}~\cite{markowitz1952portfolio} models for comparisons. We have made \textbf{4 categories} of models. Namely category \textbf{A, B, C, D}. For each category, we have at max \textbf{3 model types}, type \textbf{1, 2, 3}. The model types are described below:

\begin{itemize}
\item\textbf{Model Type-1}: Models of these types try to minimize some metric based on the risk of daily portfolio returns, while keeping the portfolio mean return greater than some specified level.

\item\textbf{Model Type-2}:
These models maximize the mean daily portfolio return penalizing some metric based on portfolio risk.

\item\textbf{Model Type-3}:
These models maximize the \textbf{Sharpe ratio}\footnote{\normalsize Throughout this paper, we will consider Annualized Sharpe Ratio\label{def:Sharpe}. There are about 252 trading days in a year. If average daily returns for a stock is $\bar{r}$, and its standard deviation is $\sigma_r$, then $$\text{\textbf{Annualized Sharpe Ratio using Daily Returns}}=\sqrt{252}\times\frac{\bar{r}}{\sigma_r}$$}~\cite{sharpe1994sharpe} of daily returns of the portfolio, penalizing some metric based on portfolio risk.
\end{itemize}

\vspace{0.2cm}
For each of the models, we want to \textbf{limit excessive short-selling} of stocks. To achieve this, we will impose some constraints. If the portfolio weight vector is defined as $w\in \mathbb{R}^d$, then we put a constraint on $\bm{||w||_1}$\footnote{\normalsize Here, $||w||_1$ denotes the $L_1$-norm for the vector $w$. For any portfolio vector $w$, $||w||_1$ is also known as the \textbf{Leverage of the portfolio}\label{def:leverage}}, which is $\bm{||w||_1 \le 2}$. It will imply that our models \textbf{will not short} more than $\bm{50\%}$ of our total capital, which can be interpreted as a form of portfolio regularization shown to improve stability and mitigate estimation error~\cite{roncalli2016regularization}. The \textbf{common constraints} that apply to all the models are:
\begin{itemize}
    \item $w^T 1 = 1$
    \item $||w||_1 \le 2$
\end{itemize}

\vspace{0.2cm}
All other \textbf{model specific constraints} are written beside the model objective functions. Hyperparameters are denoted by $\lambda_1$ \& $\lambda_2$. $\Sigma$ denotes the covariance matrix of the stock returns and $\mu$ denotes the mean vector of the stock returns. (Note, $(w^T\mu-100)$ is the percentage mean return of the portfolio \& $\bm{r}$ denotes the \textbf{minimum daily return} required in \%)

Now, we'll take a look at all the models, they are divided into different categories. The \textbf{Model Categories} differ based on the fact that how they quantify risk of portfolio returns. The naming of the models is based on their category and their type, for example, model of \textbf{Type-1} and \textbf{Category-A} will be written as model \textbf{A1}.

\noindent\textbf{Model Category-A:}  
These models use the variance of the portfolio return as a portfolio risk metric. Different model objective functions in this setup are:

\textbf{Type-1 (A1):}
\begin{equation}\label{def:A1}
    \min_w \quad w^T \Sigma w, \quad \text{subject to} \quad (w^T \mu - 100) \geq r
\end{equation}

\textbf{Type-2 (A2):}
\begin{equation}\label{def:A2}
    \max_w \quad (1-\lambda_1)(w^T \mu - 100) - \lambda_1 w^T \Sigma w
\end{equation}

\textbf{Type-3 (A3):}
\begin{equation}\label{def:A3}
    \max_w \quad \frac{(w^T \mu - 100)}{\sqrt{w^T \Sigma w}}
\end{equation}

\vspace{10pt}
\noindent\textbf{Model Category-B:}  
These models use a weighted average of portfolio variance and $\bm{d_{eq}^2}$ (subsection~\ref{subsec:square-distance}) as a risk measure. Objective functions are:

\textbf{Type-1 (B1):}
\begin{equation}\label{def:B1}
    \min_w \quad (1-\lambda_1)w^T \Sigma w + \lambda_1 d_{eq}^2(w), \quad \text{subject to} \quad (w^T \mu - 100) \geq r
\end{equation}

\textbf{Type-2 (B2):}
\begin{equation}\label{def:B2}
    \max_w \quad (1-\lambda_1-\lambda_2)(w^T \mu - 100) - \lambda_1 w^T \Sigma w - \lambda_2 d_{eq}^2(w)
\end{equation}

\textbf{Type-3 (B3):}
\begin{equation}\label{def:B3}
    \max_w \quad (1-\lambda_1)\frac{(w^T \mu - 100)}{\sqrt{w^T \Sigma w}} - \lambda_1 d_{eq}^2(w)
\end{equation}

\vspace{10pt}
\noindent\textbf{Model Category-C:}  
This category uses the correlation variance $\bm{\sigma_{\rho}^2}$ (subsection~\ref{subsec:correlation-variance}) as the risk metric. The models are:

\textbf{Type-1 (C1):}
\begin{equation}\label{def:C1}
    \min_w \quad \sigma_{\rho}^2(w), \quad \text{subject to} \quad (w^T \mu - 100) \geq r
\end{equation}

\textbf{Type-2 (C2):}
\begin{equation}\label{def:C2}
    \max_w \quad (1-\lambda_1)(w^T \mu - 100) - \lambda_1 \sigma_{\rho}^2(w)
\end{equation}

\textbf{Type-3 (C3):}
\begin{equation}\label{def:C3}
    \max_w \quad (1-\lambda_1)\frac{(w^T \mu - 100)}{\sqrt{w^T \Sigma w}} - \lambda_1 \sigma_{\rho}^2(w)
\end{equation}

\vspace{10pt}
\noindent\textbf{Model Category-D:}  
These models use a weighted average of portfolio variance and $\bm{\sigma_{\rho}^2}$. Only two types exist:

\textbf{Type-1 (D1):}
\begin{equation}\label{def:D1}
    \min_w \quad (1-\lambda_1)w^T \Sigma w + \lambda_1 \sigma_{\rho}^2(w), \quad \text{subject to} \quad (w^T \mu - 100) \geq r
\end{equation}

\textbf{Type-2 (D2):}
\begin{equation}\label{def:D2}
    \max_w \quad (1-\lambda_1-\lambda_2)(w^T \mu - 100) - \lambda_1 w^T \Sigma w - \lambda_2 \sigma_{\rho}^2(w)
\end{equation}

During the actual model training, the values of $\Sigma$, $\mu$, the $w_{eq}$ in $d^2_{eq}(w)$, matrices in $\sigma^2_{\rho}(w)$ will be replaced by their estimates. In the next part, we will see how the models are trained and the hyperparameters are tuned to make the models comparable.

\section{Methodology} \label{sec:methodology}

\subsection{Training \& Testing Period} \label{subsec:training-testing-period}

\hspace{20pt}At the beginning of each month or each quarter, our model is trained using daily return data from the previous months. We consider training periods of 6 or 12 months. The portfolio constructed by the model is then deployed for a subsequent period of 1 months. We refer to the first duration as the \textbf{training period} (in months) and the latter as the \textbf{testing period} (in months). Since we don't want to change the portfolio too frequently, at the start of each month, we would calculate optimal portfolio vector based on our models and use that portfolio vector throughout that month. At the end of each month, we will store the value of monthly \% returns for that portfolio.

\subsection{Stock Selection Method} \label{subsec:stock-selection-method}

\hspace{20pt}Based on the daily returns from the training period, we construct a correlation matrix including all of the stock returns. This matrix captures the \textbf{Spearman correlation}~\cite{daniel1990applied} between each pair of stock returns. We have used Spearman correlation, since it's a robust measure of correlation between two returns series. Our objective is to identify groups of stocks where \textbf{within-group} correlations are \textbf{high}, while \textbf{between-group} correlations remain \textbf{low}. To achieve this, we employ \textbf{hierarchical clustering}~\cite{nielsen2016hierarchical}. The distance between any two stocks is defined as $(1 - r)$, where $r$ represents the Spearman correlation between their daily returns. A smaller distance indicates a stronger positive correlation, which makes the stocks more likely to belong to the same group. We cluster the stocks using hierarchical clustering using this distance matrix.

\begin{figure}[H]
    \centering
    \includegraphics[width = \linewidth]{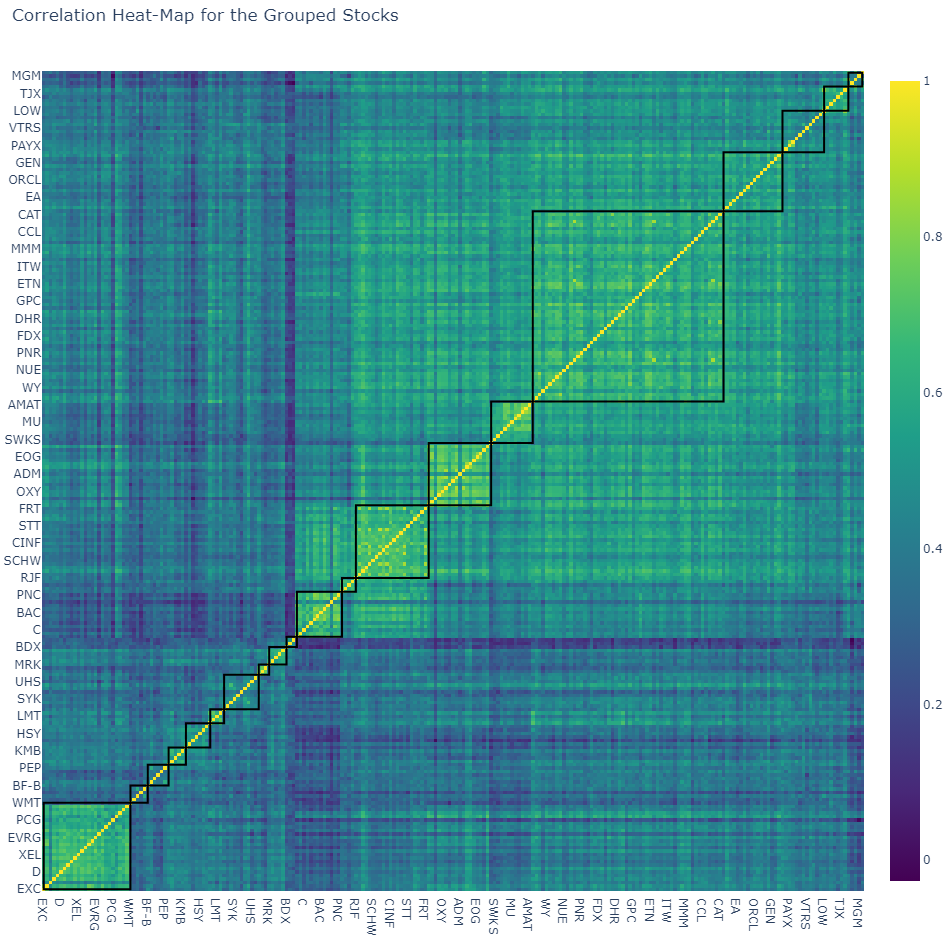}
    \caption{Correlation Matrix Heatmap of 350 stocks \& Forming 20 Cluster of Stocks (Shown in black boxes)}
    \label{fig:corr_heat_map}
\end{figure}

Once the groups are formed, from each group, we select the stock with the highest \textbf{Sharpe ratio} (definition \ref{def:Sharpe}) based on the training period daily returns. In the \textbf{heatmap} (figure \ref{fig:corr_heat_map}) of the Correlation matrix (symmetric). Along the horizontal and vertical axes, we have the ticker of the stocks. The black boxes show the stock clusters. Each cell color represents the correlation between a pair of stocks. So, for a particular training period, for each model, the selected stock remains the same, since selected stocks only depend on the estimation period and all the models create portfolio on the same set of stocks.

\subsection{Covariance Matrix \& Mean Estimate} \label{subsec:covariance-matrix-mean-estimate}

\hspace{20pt}After selecting the stocks, we will need estimates of mean return of each stock returns and also estimate of the covariance matrix for the selected stock returns. Whenever working with data on stock returns, we have to be careful about the estimate of the covariance matrix of stock returns based on historical data, since the estimate can be very \textbf{unreliable}, those problems are well documented by Jobson and Korkie~\cite{jobson1980estimation}. So, we use two \textbf{shrinkage estimates}~\cite{boas2017shrinkage} namely \textbf{Unbiased Shrinkage Correlation Covariance}\label{def:USCC} (\textbf{USCC})~\cite{Kwan2017} \& \textbf{Shrinkage Covariance}\label{def:SC} (\textbf{SC})~\cite{Ledoit2004}. Both estimates provide some weighted average of \textbf{sample covariance matrix} ($\hat{\Sigma}$) and \textbf{constant correlation covariance matrix} (F), where in F we assume the \textbf{correlation between any two stocks are same}. The shrinkage estimate S then becomes,

$$
S = (1 - \lambda)\hat{\Sigma}+\lambda F, \quad (\lambda \in (0,1))
$$
\noindent Those two methods provide two different methods to estimate optimal value of $\lambda$. For the estimate of mean stock returns, we just take the mean of the daily stock returns based on the training period data.

\subsection{Hyperparameter Tuning} \label{subsec:hyperparameter-tuning}

\hspace{20pt}Since most of our models include hyperparameters, we require a systematic approach to tune them~\cite{bergstra2012random}, ensuring model comparability. In the models we have constructed, hyperparameters take values within the range $[0,1]$. For example, consider a model with a 12-month training period and a 1-month testing period for the year 2013.

To determine the optimal hyperparameter values for a particular model, we first run the model on the year of 2012 using different hyperparameter settings. The performance of each configuration is evaluated based on approximately 252 daily returns obtained during the year 2012. An evaluation function assesses these returns, and the optimal hyperparameter value is selected. This chosen value is then used throughout 2013. Following sections outline detailed methodology used for both single and two-hyperparameter cases.

\subsubsection{1-Hyperparameter Case} \label{subsubsec:1-hyperparameter-case}

\hspace{20pt}When we have a single hyperparameter say $\lambda_1$, its values are selected from the set $\{0.01,0.02,\dots,\\0.99\}$. Based on these hyperparameter values, we optimize the objective function using daily returns from the training period. From the resulting portfolio vector, we compute a corresponding vector of daily portfolio returns. In our example case, we will have 252 daily returns corresponding to each model and hyperparameter value. To evaluate the model, we define a \textbf{model evaluation value} as:
$$
-\frac{\text{ES}}{\text{Sharpe}}
$$
Let \(\tilde{r}\) represent the vector of portfolio return values. Then \textbf{Expected Shortfall (ES)}\label{def:ES}~\cite{rockafellar2000optimization} is defined as the mean of all returns \( r_i \) in \(\tilde{r}\) that satisfy \( r_i < 0 \):
$$
ES = \text{mean}(r_i \mid r_i < 0)
$$
Since, we would like to reduce the value of -(ES) and increase the value of Sharpe, we have defined the model evaluation value like that. A \textbf{smaller model evaluation value indicates a better model}.

We now have a vector of length \textbf{99} containing model evaluation values. To reduce \textbf{noise} in hyperparameter tuning, we apply a \textbf{mean filter}~\cite{al-amri2010comparative}: a \textbf{slider} of length \textbf{11} moves across the vector, replacing each value with the \textbf{average} of its nearest 11 points (including itself). Finally, from this filtered vector, we select the optimal value $\bm{\lambda_1^*}$, corresponding to the lowest value in the filtered vector.

\subsubsection{2-Hyperparameter Case} \label{subsubsec:2-hyperparameter-case}

\begin{figure}[ht]
    \centering
    \includegraphics[width = 0.75\linewidth]{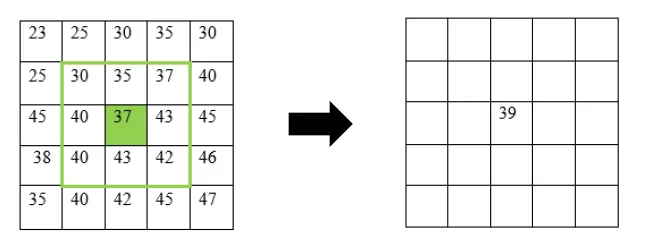}
    \caption{Mean Filter for a matrix with kernel matrix of size 3}
    \label{fig:mean_filter}
\end{figure}

\hspace{20pt}When we have two hyperparameters $\lambda_1$ \& $\lambda_2$, their values are selected from the set $\{0.01,0.02,\\\dots,0.99\}$. But, here we need to add one additional constraint $\bm{\lambda_1+\lambda_2 \le 1}$. Based on these hyperparameter values, we optimize the objective function using daily returns from the training period. From the resulting portfolio vector, we compute a corresponding vector of daily portfolio returns for the training period. To evaluate the model, we use the same \textbf{model evaluation value} as before, which is
$$
(-\frac{\text{ES}}{\text{Sharpe}})
$$
Now, we have a $99\times99$ matrix of model evaluation values, where the \textbf{lower right triangle} has all values \textbf{zero} due to the constraint $\lambda_1+\lambda_2 \le 1$. Here we also apply \textbf{mean filter}(figure \ref{fig:mean_filter}) with the window size 11 (which means that the kernel matrix is of shape $11\times11$).

From the filtered matrix, we select the optimal values $\bm{\lambda_1^*}$ \& $\bm{\lambda_2^*}$, corresponding to the lowest values in that matrix.

\subsection{Methodology For Model Testing} \label{subsec:methodology-model-testing}

\hspace{20pt}Here, we look at the methodology which we'll apply to optimize portfolio based on each model. We will generate optimal portfolio weight vector and daily portfolio returns for each month.

\begin{algorithm}[H]
\caption{Annual Hyperparameter Tuning and Monthly Portfolio Optimization}
\label{alg:hyperparam-portfolio}
\setstretch{1.2}
\begin{algorithmic}[1]
    \Require Daily returns time series $\mathcal{R}$; training window length $W$ months (6 or 12 months)
    \Ensure Portfolio allocation vector $w_{Y,m}$ \& daily portfolio return $r_{Y,m}$ for each month $m$ in year $Y$
    
    \State Determine range of years in $\mathcal{R}$: $Y_{\min} (2010)$ to $Y_{\max}(2019)$
    
    \For{year $Y \gets Y_{\min}$ to $Y_{\max}$}
        \Comment{Tune hyperparameters at the start of year $Y$}
        \State Collect daily returns from year $Y-1$
        \State Perform hyperparameter tuning on data from year $Y-1$
        \State Store optimized hyperparameters $\Theta_Y$
        
        \For{month $m \gets 1$ to $12$}
            \Comment{Run portfolio model for month $m$}
            \State Define testing period $T_{\text{test}}$ as month $m$ of year $Y$
            \State Define training period $T_{\text{train}}$ as the $W$ months before $T_{\text{test}}$
            \State Extract daily returns data from $\mathcal{R}$ over $T_{\text{train}}$
            \State Select stocks based on data in $T_{\text{train}}$
            \State Estimate mean return vector $\mu$ and covariance matrix $\Sigma$ over $T_{\text{train}}$
            \State Retrieve hyperparameters $\Theta_Y$
            \State Solve portfolio optimization model with:
            \Statex \hspace{50pt} Selected stocks, $\mu$, $\Sigma$, and $\Theta_Y$
            \State Obtain portfolio vector $w_{Y,m}$ for month $m$ of year $Y$
            \State Obtain daily portfolio return vector $r_{Y,m}$ for month $m$ of year $Y$ 
            \State Store $w_{Y,m}$ and $r_{Y,m}$
        \EndFor
    \EndFor
    
    \Return All monthly portfolio vectors $\{w_{Y,m}\}$ \& portfolio return vectors $\{r_{Y,m}\}$
\end{algorithmic}
\end{algorithm}
\vspace{-0.3cm}

For cases where this objective function is convex, we employ the \textbf{MOSEK} solver~\cite{mosek2016cookbook} to minimize it and determine the optimal portfolio weight vector and in non-convex scenarios, the optimization is performed using the \textbf{trust-constr}~\cite{conn2000trust} algorithm, applied to \textbf{daily returns from the training period}. This optimization process yields the optimal asset allocation across the selected stocks. The resulting \textbf{portfolio} is then \textbf{held} unchanged over the \textbf{testing period}. Thus we get the portfolio weight vectors and portfolio daily returns, from which then we calculate model evaluation metrics discussed in the next section.

\subsection{Methodology For Model Evaluation} \label{subsec:methodology-model-evaluation}

\hspace{20pt}After running the model for 10 years from 2010-2019, we have 120 \textbf{monthly returns} for each portfolio models. For each \textbf{(training period, testing period)} pair and corresponding to each model type, we will look at mean monthly return of the portfolio, standard deviation and Sharpe of portfolio monthly returns. Since we have \textbf{monthly returns} here, to annualize the Sharpe ratio, we will multiply by $\sqrt{12}$ and not by $\sqrt{252}$ in footnote \ref{def:Sharpe}. \textbf{Expected Shortfall} is calculated as defined in sub-subsection \ref{subsubsec:1-hyperparameter-case}, the same definition holds for monthly returns. We'll also look at \textbf{mean leverage} (footnote \ref{def:leverage}) of the portfolio.

Apart from these measures, we will look at some measures based on \textbf{daily returns} of the portfolio. We'll take a look at \textbf{Sharpe Ratio} of the Portfolio \textbf{daily returns}. We'll also take a look at \textbf{drawdown}\label{def:drawdown} of the portfolio. Let \( P_t \) be the cumulative return of the portfolio at time \( t \), and let \( M_t \) be the running maximum of the cumulative return up to time \( t \), defined as $M_t = \max_{s \leq t} R_s$. Then, the drawdown at time \( t \), denoted \( D_t \), is defined as:
$$
D_t = \frac{P_t - M_t}{M_t}
$$
\hspace{20pt}Thus for each portfolio model, we will have a series of drawdown values for each day throughout the years 2010-19. As performance measure we will take mean of the drawdown series of the portfolio. \textbf{Mean daily drawdown} is always non-positive, as it represents the average decline from the historical peak of the portfolio value. A \textbf{less negative} (i.e., closer to zero) mean drawdown indicates \textbf{lower downside risk} and suggests a more stable and potentially better-performing portfolio. Also, we will look at \textbf{Sharpe Ratio} of \textbf{weekly returns} of the portfolio. For calculating Sharpe Ratio of the weekly returns, we will use weekly returns and use the multiplier $\sqrt{52}$ instead of $\sqrt{252}$ in footnote \ref{def:Sharpe} (Since we have about 52 weeks in a year).

\section{Portfolio Optimization Results} \label{sec:portfolio-optimization-results}

\subsection{Stock Universe Overview} \label{subsec:stock-universe}

\begin{figure}[ht]
    \centering
    \includegraphics[width = 0.81\linewidth]{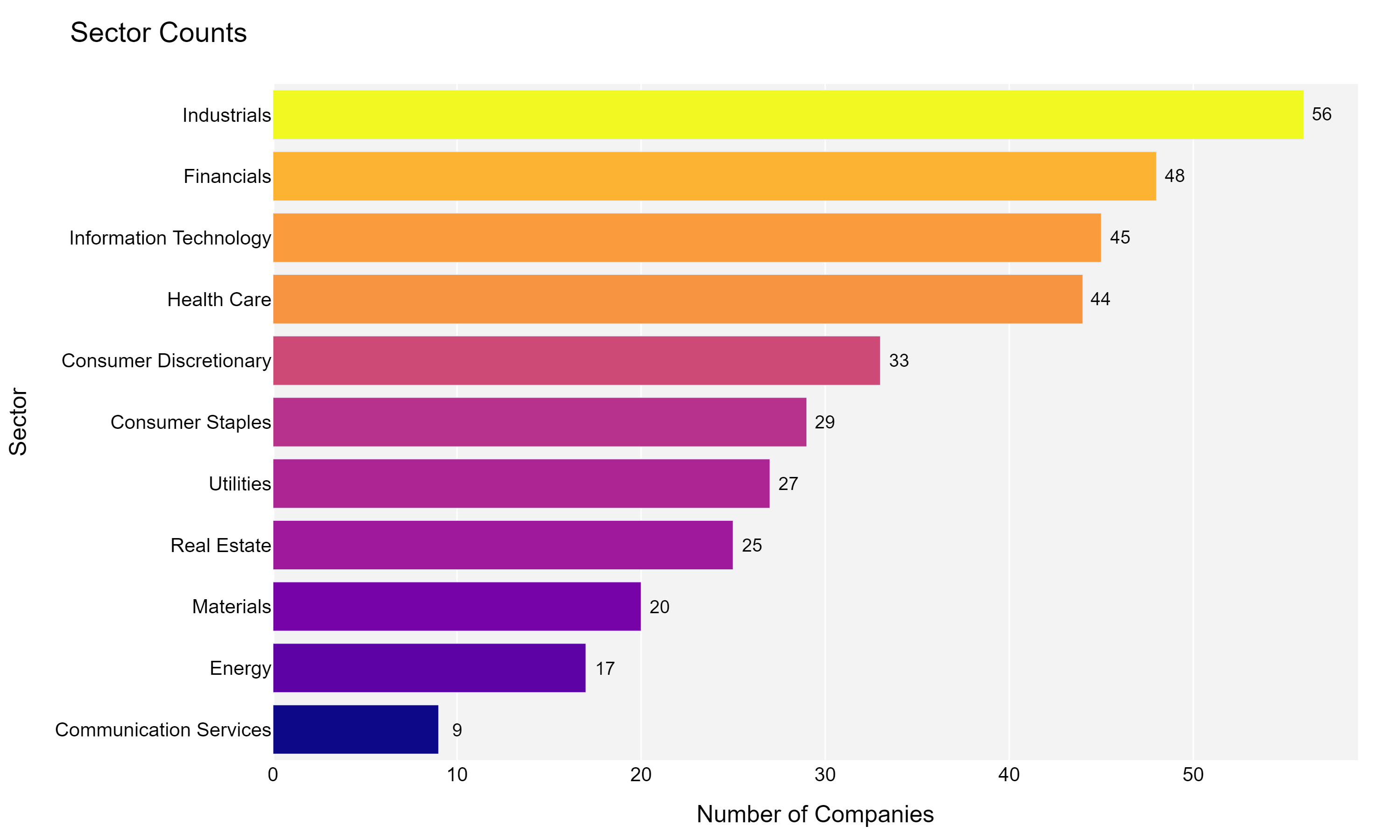}
    \caption{Bar Chart of Sectors of The Stocks of Our Universe}
    \label{fig:sector_bar}
\end{figure}

Before taking a look at the results, we will take a look at the \textbf{stocks we are working with}. From Figure~\ref{fig:sector_bar} Our \textbf{universe} consists of about \textbf{350 stocks}, which are part of the S\&P500. First we take a look at the Sectors of the stocks in our universe. The stock universe is predominantly composed of companies from the \textbf{Industrials} (56), \textbf{Financials} (48), and \textbf{Information Technology} (45) sectors, which form the core of the portfolio. The \textbf{Health Care} sector also represents a significant share with 44 companies. Other notable sectors include \textbf{Consumer Discretionary} (33) and \textbf{Consumer Staples} (29). Conversely, \textbf{Communication Services} has the fewest companies with only 9, followed by \textbf{Energy} (17) and \textbf{Materials} (20). This distribution highlights a strong concentration in industrial and technology sectors, with smaller representation from energy and materials industries.

\begin{figure}[H]
    \centering
    \begin{subfigure}[b]{0.48\linewidth}
        \centering
        \includegraphics[width=\linewidth]{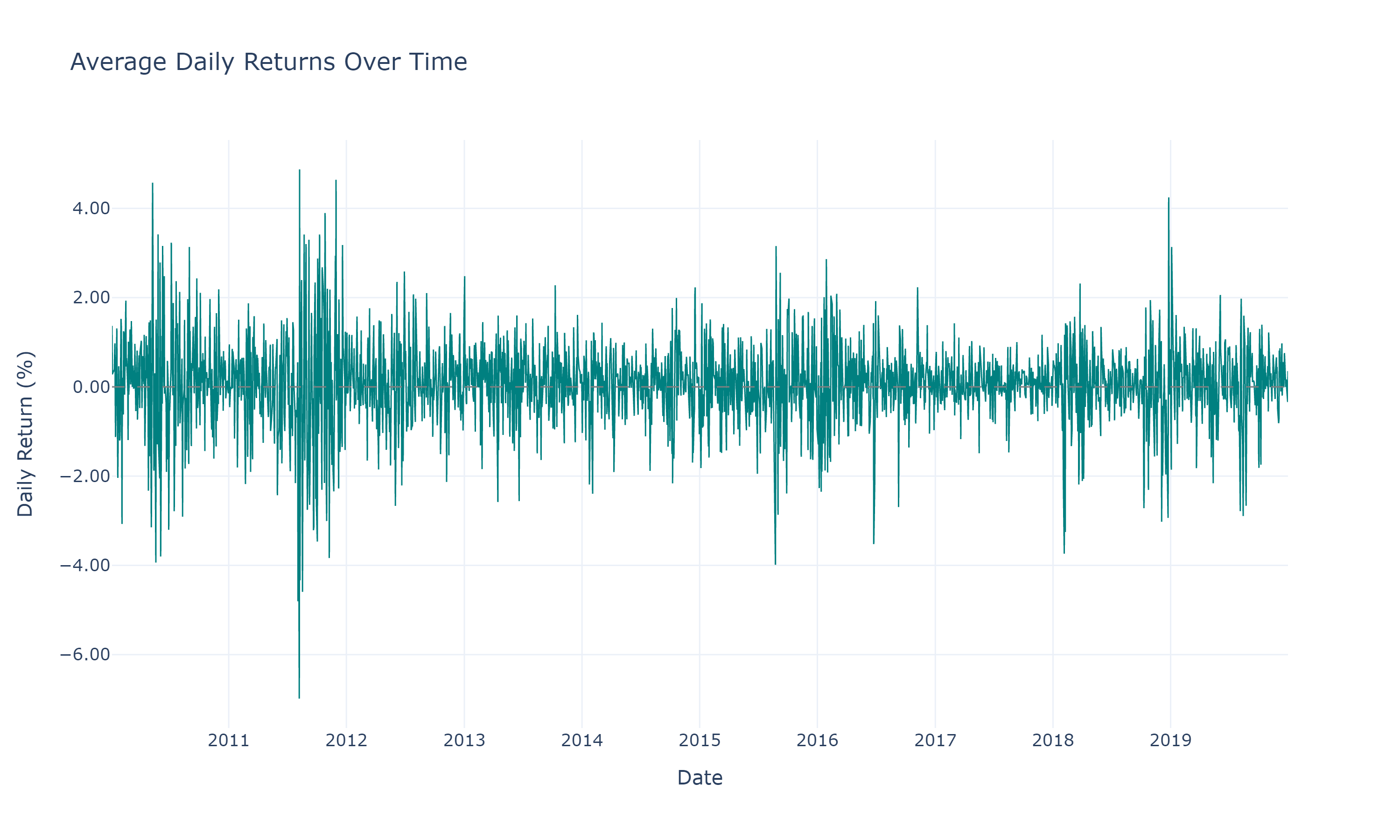}
        \caption{Mean Daily Returns}
        \label{fig:uni_1}
    \end{subfigure}
    \hfill
    \begin{subfigure}[b]{0.48\linewidth}
        \centering
        \includegraphics[width=\linewidth]{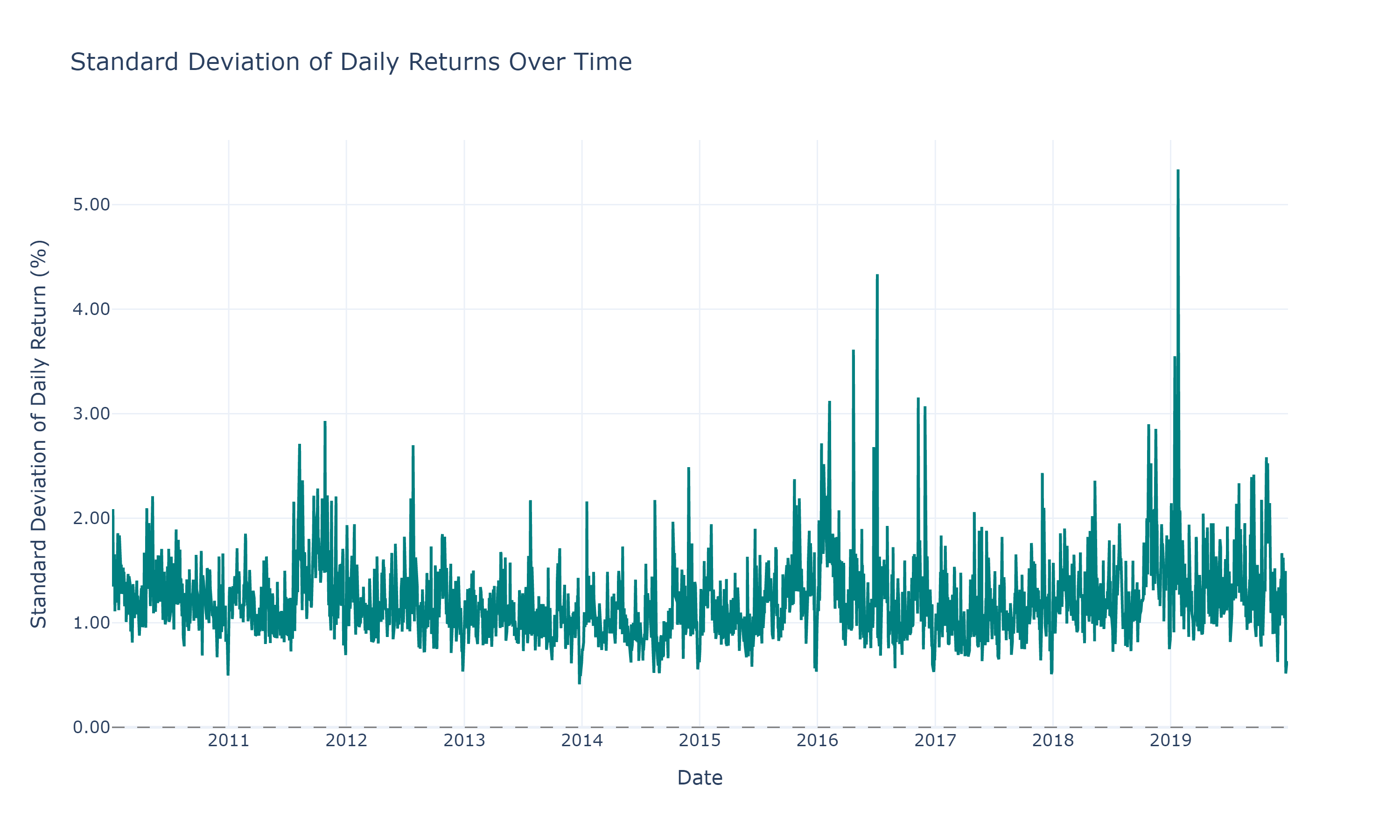}
        \caption{Daily Returns Standard Deviation}
        \label{fig:uni_2}
    \end{subfigure}
    
    \vspace{0.5cm} 

    \begin{subfigure}[b]{0.48\linewidth}
        \centering
        \includegraphics[width=\linewidth]{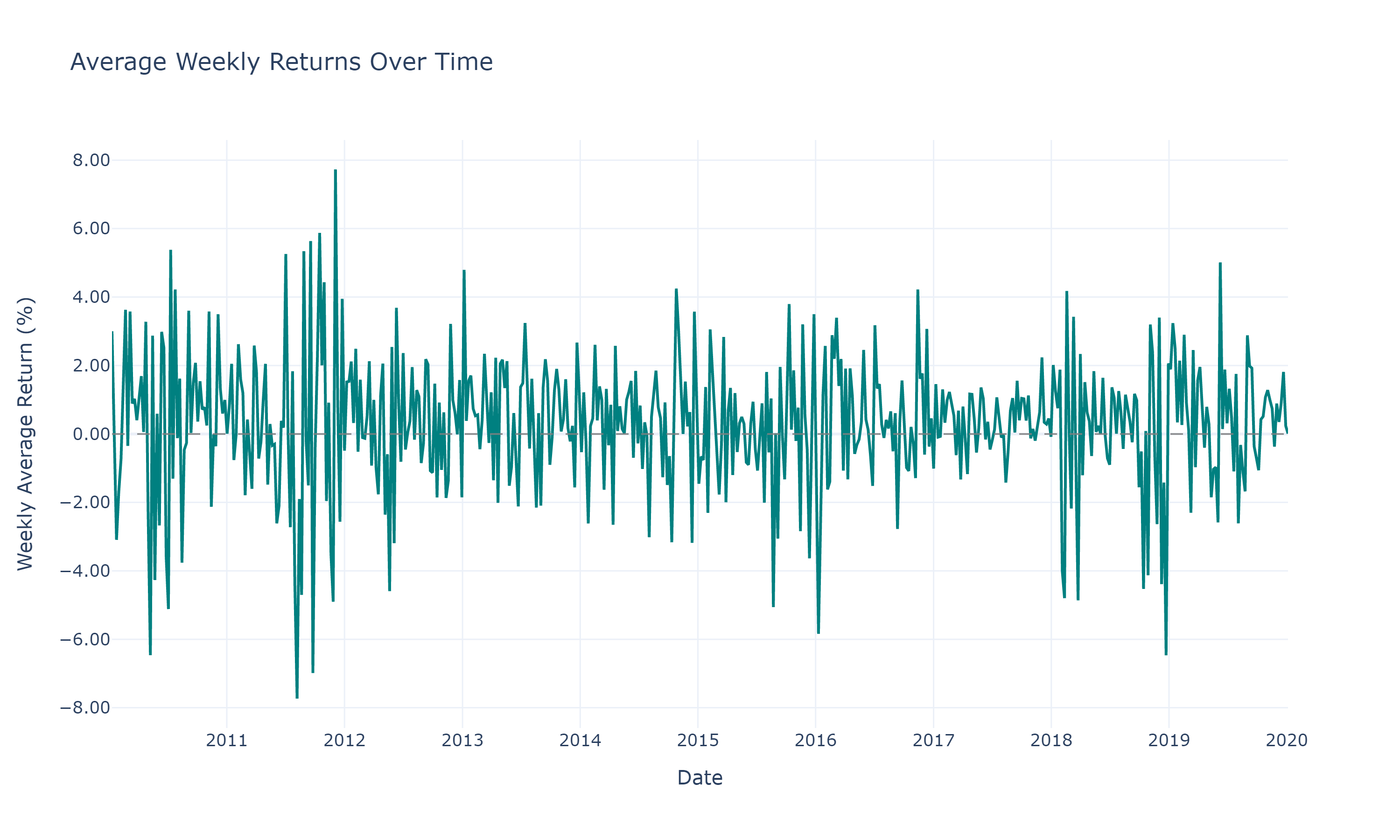}
        \caption{Mean Weekly Returns}
        \label{fig:uni_3}
    \end{subfigure}
    \hfill
    \begin{subfigure}[b]{0.48\linewidth}
        \centering
        \includegraphics[width=\linewidth]{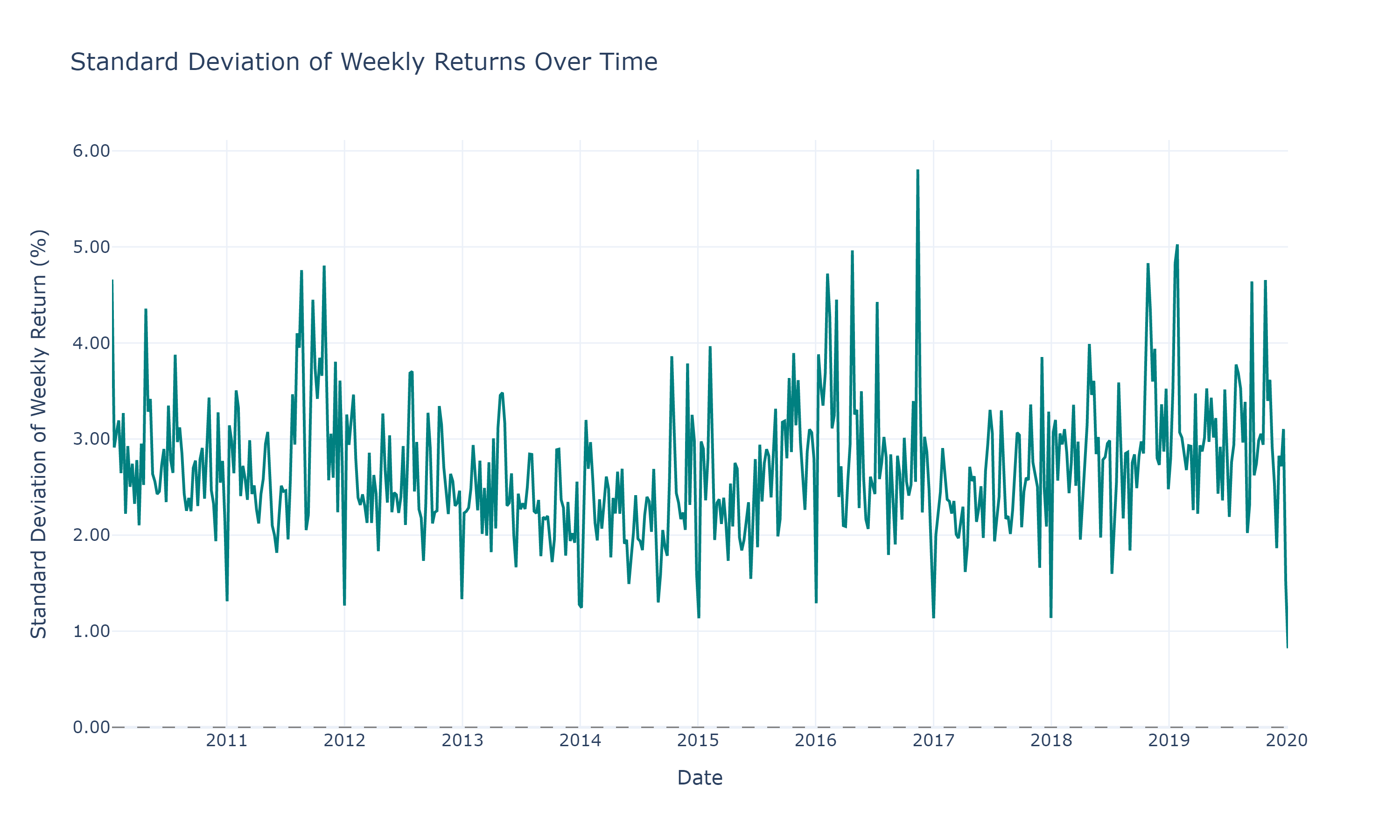}
        \caption{Weekly Returns Standard Deviation}
        \label{fig:uni_4}
    \end{subfigure}
    
    \vspace{0.5cm} 

    \begin{subfigure}[b]{0.48\linewidth}
        \centering
        \includegraphics[width=\linewidth]{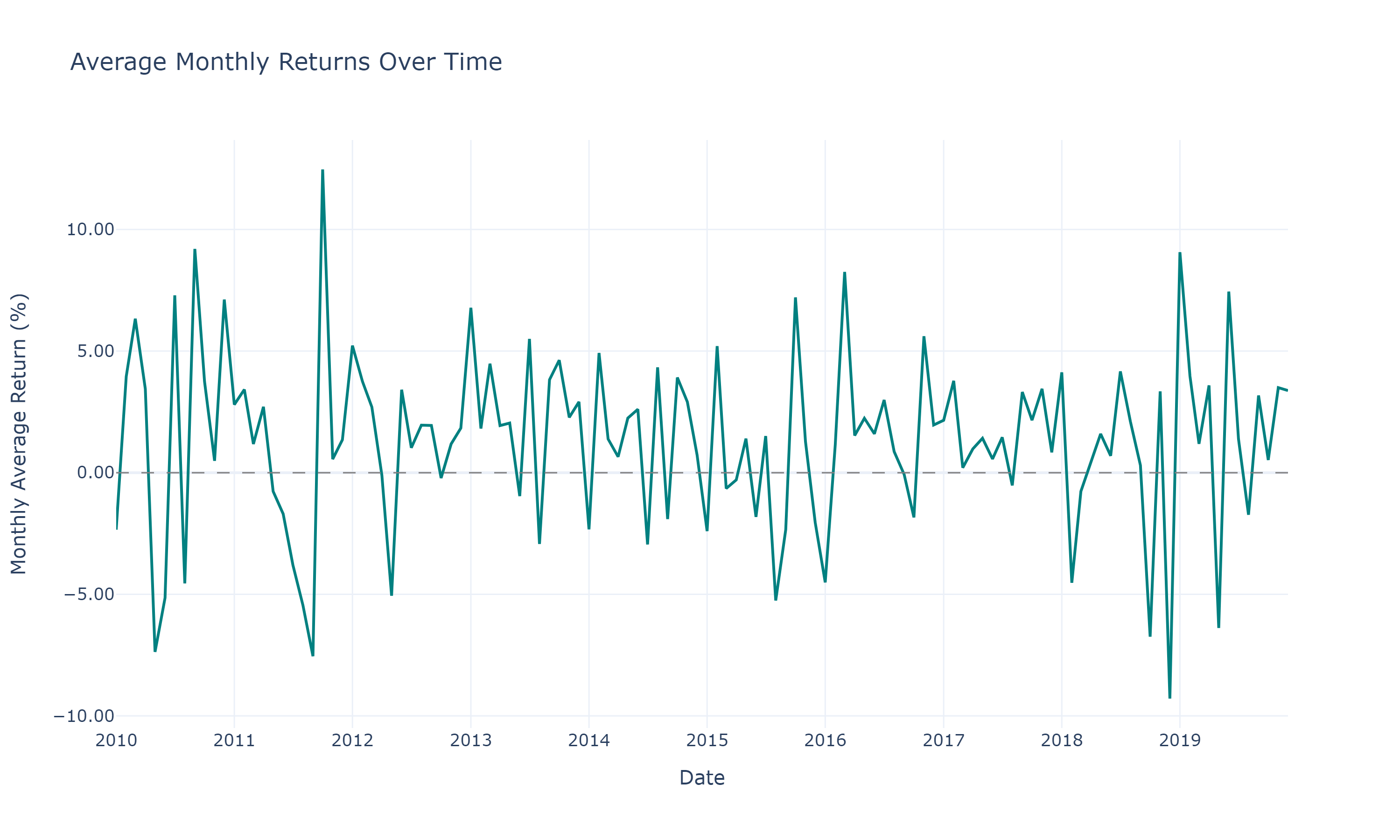}
        \caption{Mean Monthly Returns}
        \label{fig:uni_5}
    \end{subfigure}
    \hfill
    \begin{subfigure}[b]{0.48\linewidth}
        \centering
        \includegraphics[width=\linewidth]{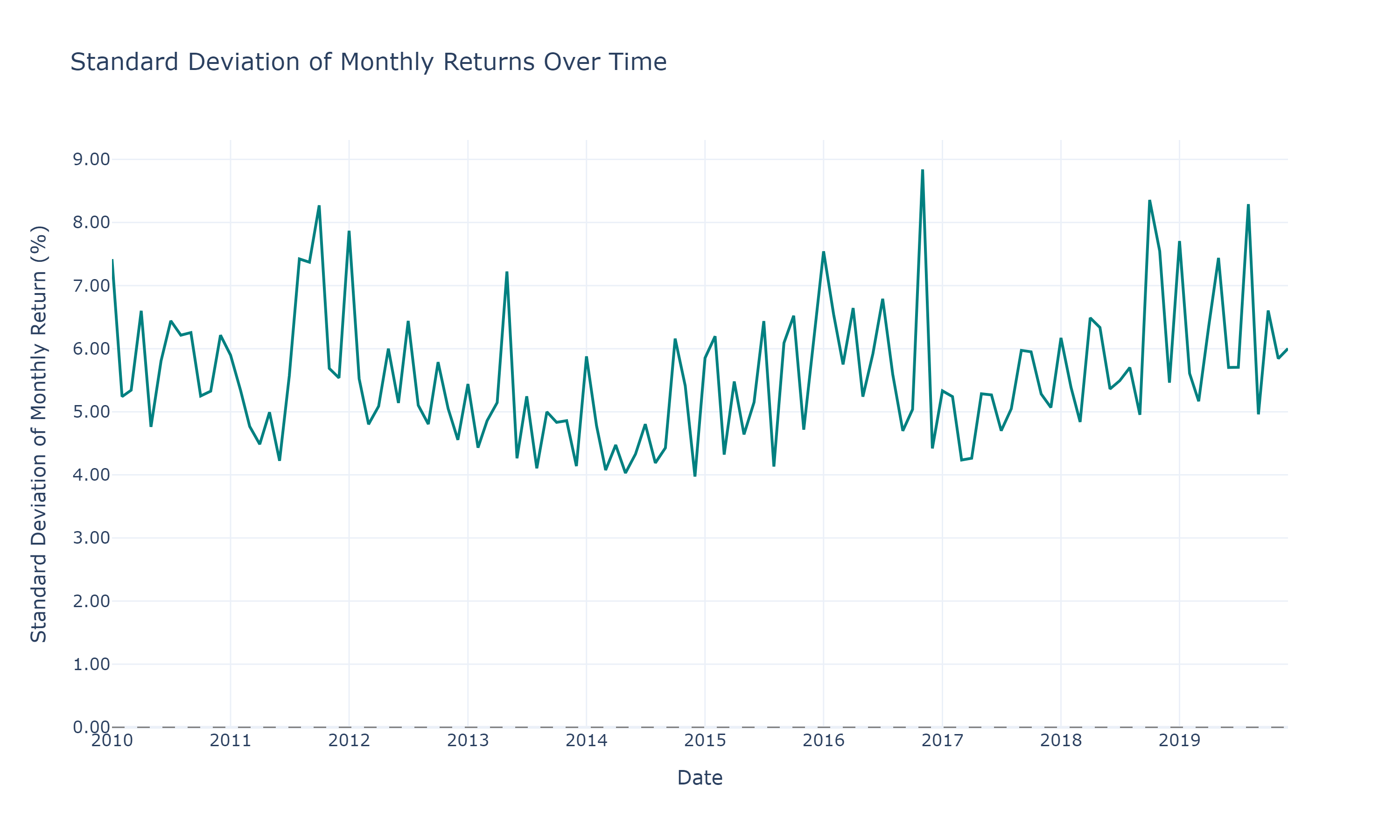}
        \caption{Monthly Returns Standard Deviation}
        \label{fig:uni_6}
    \end{subfigure}

    \caption{Overview of Return Statistics at Daily, Weekly, and Monthly Frequencies}
    \label{fig:universe_summary}
\end{figure}

In Figure \ref{fig:universe_summary} we will look at the mean returns of all the stocks in the universe. Basically, for each day during the years 2010-19, we have calculated the mean \% return of all the stocks in the universe. We'll also calculated standard deviation of the daily returns. For any particular day, we look at the returns of all the stocks and calculate their standard deviation and plot them along time. We did the same thing also for weekly and monthly returns for all the stocks in our universe. Now, we move on to the results of our Portfolio Optimization models.

\subsection{Training \& Testing Period (12,1)} \label{subsec:training-testing-period-(12,1)}

\hspace{20pt}For the case where the training period is 12 months and the testing period is 1 months, we will examine the evaluation charts for the models, and look at some descriptive statistics. As mentioned earlier in subsection \ref{subsec:stock-selection-method}, for a particular training period \& testing period pair, stock selection only depends on the training period. Hence, all models train on the same set of stocks for any particular testing month. And as the testing month changes, so does the set of selected stocks, since the training data changes as we move from one month to another month. For this case (12,1), for the 120 months, \textbf{Top-20 most selected stocks} are given in the table. The columns represent the name of the stock, the \textbf{Sub-Industry} of the stock and number of months the stock was part of the portfolio.

\begin{table}[H]
\centering
\caption{(12,1) Period Top 20 Stock Selection Summary}
\label{tab:(12,1)-stock_selection}
\resizebox{\textwidth}{!}{%
  \setlength{\extrarowheight}{2pt}%
  \renewcommand{\arraystretch}{1.1}%
  \begin{tabular}{|c|c|c|}
    \toprule
    \hline
    \rowcolor{teal!70}
    \color{white}\textbf{Stock Name}
      & \color{white}\textbf{Sub-Industry}
      & \color{white}\textbf{Months Selected} \\
    \hline
    Exxon Mobil
      & Integrated Oil \& Gas
      & 112 \\ \hline
    Johnson \& Johnson
      & Pharmaceuticals
      & 82  \\ \hline
    UnitedHealth Group
      & Managed Health Care
      & 81  \\ \hline
    NVR, Inc.
      & Homebuilding
      & 76  \\ \hline
    Waste Management
      & Environmental \& Facilities Services
      & 69  \\ \hline
    McDonald's
      & Restaurants
      & 60  \\ \hline
    Southern Company
      & Electric Utilities
      & 59  \\ \hline
    Walmart
      & Consumer Staples Merchandise Retail
      & 58  \\ \hline
    U.S. Bancorp
      & Diversified Banks
      & 54  \\ \hline
    Ecolab
      & Specialty Chemicals
      & 52  \\ \hline
    Home Depot (The)
      & Home Improvement Retail
      & 49  \\ \hline
    Federal Realty Investment Trust
      & Retail REITs
      & 46  \\ \hline
    IBM
      & IT Consulting \& Other Services
      & 43  \\ \hline
    Lockheed Martin
      & Aerospace \& Defense
      & 42  \\ \hline
    Coca-Cola Company (The)
      & Soft Drinks \& Non-alcoholic Beverages
      & 38  \\ \hline
    W. R. Berkley Corporation
      & Property \& Casualty Insurance
      & 38  \\ \hline
    Intel
      & Semiconductors
      & 38  \\ \hline
    Costco
      & Consumer Staples Merchandise Retail
      & 35  \\ \hline
    3M
      & Industrial Conglomerates
      & 35  \\ \hline
    Texas Instruments
      & Semiconductors
      & 31  \\
    \hline
    \bottomrule
  \end{tabular}%
}
\end{table}

From Table \ref{tab:(12,1)-stock_selection} we can see, our stock selection procedure selected \textbf{high-quality}, \textbf{well-established} companies across diverse sectors, emphasizing stability, dividends, and resilience. The portfolio includes industry leaders (Exxon Mobil, Johnson \& Johnson, McDonald's) and defensive stocks (Coca-Cola, Waste Management) that perform well in downturns. This aligns with prior findings that defensive sectors tend to offer downside protection during market crises \cite{bartram2009noplace}. However, it lacks high-growth tech stocks (e.g., Apple, NVIDIA) and small/mid-cap exposure, which may limit future upside. Overall, it's a strong value-focused selection. Now, we will look into details of individual models.

\subsubsection{Type-1 Models} \label{subsubsec:(12,1)-type-1-models}

\hspace{20pt}Here, we have trained our models to have a minimum monthly return of 3\%. If we consider a month has \textbf{20 trading days}, to achieve a \textbf{3\%} \textbf{monthly return}, we will need to have a \textbf{daily return} of about \textbf{0.148\%}\footnote{Here we have used this formula of compound interest $$\text{daily return(\%)} = 100\times((1+\frac{\text{monthly return(\%)}}{100})^{\frac{1}{20}} - 1)$$}. We have used this value of daily return in each of the objective functions of our type-1 models. Now we'll look at different performance metrics of our models. These metrics were defined earlier in subsection \ref{subsec:methodology-model-evaluation}. To understand the naming here, model name 'A1-USCC' represents model of Type-1, Category-A and the estimate of covariance matrix used is USCC.

\begin{table}[H]
\resizebox{\textwidth}{!}{%
{
    \setlength{\extrarowheight}{2pt}%
    \renewcommand{\arraystretch}{1.2}%
    \begin{tabular}{|l|ccc|cc|ccc|}
    \toprule
    \hline
    \multirow{2}{*}{\shortstack[c]{\textbf{Model}}} & 
    \multirow{2}{*}{\shortstack[c]{\textbf{Mean Monthly}\\\textbf{Return(\%)}}} & 
    \multirow{2}{*}{\shortstack[c]{\textbf{Monthly Return}\\\textbf{Std(\%)}}} & 
    \multirow{2}{*}{\shortstack[c]{\textbf{Monthly}\\\textbf{ES}}} & 
    \multirow{2}{*}{\shortstack[c]{\textbf{Mean}\\\textbf{Leverage}}} & 
    \multirow{2}{*}{\shortstack[c]{\textbf{Mean Daily}\\\textbf{Drawdown(\%)}}} & 
    \multicolumn{3}{|c|}{\textbf{Sharpe}} \\
    \cline{7-9}
    & & & & & & 
    \shortstack{\textbf{Daily}} & 
    \shortstack{\textbf{Weekly}} & 
    \shortstack{\textbf{Monthly}} \\
    \hline
    \large A1-SC   & \large 12.907 & \large 12.821 & \large -3.068 & \large 1.894 & \large -3.819 & \large 0.974 & \large 2.087 & \large 1.007 \\
    \large A1-USCC & \large 13.406 & \large 12.737 & \large -2.991 & \large 1.905 & \large -3.723 & \large 1.016 & \large 2.182 & \large 1.053 \\
    \large B1-SC   & \large 12.139 & \large 12.908 & \large -2.891 & \large \textbf{1.766} & \large -4.606 & \large 0.884 & \large 1.949 & \large 0.940 \\
    \large B1-USCC & \large 12.139 & \large 13.033 & \large -2.819 & \large \textbf{1.806} & \large -4.954 & \large 0.875 & \large 1.931 & \large 0.931 \\
    \large C1-SC   & \large \textbf{17.407} & \large 12.215 & \large \textbf{-2.618} & \large 1.844 & \large \textbf{-3.455} & \large \textbf{1.168} & \large \textbf{2.954} & \large \textbf{1.425} \\
    \large C1-USCC & \large \textbf{17.999} & \large \textbf{12.083} & \large \textbf{-2.367} & \large 1.930 & \large \textbf{-3.156} & \large \textbf{1.218} & \large \textbf{3.088} & \large \textbf{1.490} \\
    \large D1-SC   & \large 14.519 & \large 12.227 & \large -2.903 & \large 1.875 & \large -3.652 & \large 1.078 & \large 2.461 & \large 1.187 \\
    \large D1-USCC & \large 14.788 & \large \textbf{12.203} & \large -2.683 & \large 1.882 & \large -3.610 & \large 1.091 & \large 2.512 & \large 1.212 \\
    \hline
    \bottomrule
    \end{tabular}%
}}
\caption{Performance Metrics Table For Type 1 Models}
\label{tab:(12,1)_type_1_summary_stat}
\end{table}
\vspace{-0.6cm}

Table~\ref{tab:(12,1)_type_1_summary_stat} uses monthly, daily and weekly returns of the portfolio. In this table from left to right the first column represent the model name, the second column represents \textbf{annualized mean monthly return(\%)}\footnote{\normalsize Let mean monthly return be $\bar{R}_M$, then annualized mean return ($\bar{R}_A$) is $12\times\bar{R}_M$.} from the model, next column has \textbf{annualized standard deviation(\%)}\footnote{\normalsize Let standard deviation of monthly returns be $\sigma_M$, then annualized standard deviation($\sigma_A$) is $\sqrt{12}\times\sigma_M$} of monthly returns. Subsequent columns contain \textbf{expected shortfall} of monthly returns, \textbf{mean leverage} and \textbf{mean daily drawdown(\%)}. Last three columns contain \textbf{Sharpe Ratio} of each portfolio based on daily, weekly and monthly returns. All these metrics are defined in subsection \ref{subsec:methodology-model-evaluation}.

We can see that, \textbf{C1(\ref{def:C1}) models} (Category-C type 1 model and its objective function is given in equation~\ref{def:C1}) perform the best in terms of mean monthly return, Monthly Expected Shortfall, Mean Daily Drawdown and Sharpe ratio. \textbf{D1(\ref{def:D1}) models} are the second-best performers, slightly trailing C1 models but still outperforming A1 and B1 models. \textbf{A1(\ref{def:A1}) and B1(\ref{def:B1}) models} exhibit comparable performance—one slightly excels in the Sharpe ratio, while the other performs better in ES. However, both under-perform relative to C1 and D1 models. \textbf{B1} models show slight \textbf{lower leverage} than other models. Using \textbf{USCC} estimate for the covariance matrix yields significantly better results compared to \textbf{SC} estimate. These estimates are defined in subsection \ref{subsec:covariance-matrix-mean-estimate}.

\subsubsection{Type-2 Models} \label{subsubsec:(12,1)-type-2-models}

\hspace{20pt}Now, we take a look at Type-2 models. We have models of 4 categories (A, B, C \& D) under Type-2 models. Here also we take a look at the performance of different models.

\begin{table}[H]
\resizebox{\textwidth}{!}{%
{
    \setlength{\extrarowheight}{2pt}%
    \renewcommand{\arraystretch}{1.2}%
    \begin{tabular}{|l|ccc|cc|ccc|}
    \toprule
    \hline
    \multirow{2}{*}{\shortstack[c]{\textbf{Model}}} & 
    \multirow{2}{*}{\shortstack[c]{\textbf{Mean Monthly}\\\textbf{Return(\%)}}} & 
    \multirow{2}{*}{\shortstack[c]{\textbf{Monthly Return}\\\textbf{Std(\%)}}} & 
    \multirow{2}{*}{\shortstack[c]{\textbf{Monthly}\\\textbf{ES}}} & 
    \multirow{2}{*}{\shortstack[c]{\textbf{Mean}\\\textbf{Leverage}}} & 
    \multirow{2}{*}{\shortstack[c]{\textbf{Mean Daily}\\\textbf{Drawdown(\%)}}} & 
    \multicolumn{3}{|c|}{\textbf{Sharpe}} \\
    \cline{7-9}
    & & & & & & 
    \shortstack{\textbf{Daily}} & 
    \shortstack{\textbf{Weekly}} & 
    \shortstack{\textbf{Monthly}} \\
    \hline
    \large A2-SC   & \large 12.264 & \large 14.190 & \large -3.006 & \large 1.728 & \large -7.167 & \large 0.637 & \large 1.792 & \large 0.864 \\
    \large A2-USCC & \large \textbf{14.673} & \large 13.843 & \large -2.837 & \large 1.699 & \large -3.831 & \large 0.818 & \large 2.197 & \large 1.060 \\
    \large B2-SC   & \large 13.759 & \large \textbf{10.223} & \large \textbf{-2.213} & \large \textbf{1.392} & \large \textbf{-2.626} & \large \textbf{1.227} & \large \textbf{2.790} & \large \textbf{1.346} \\
    \large B2-USCC & \large 14.341 & \large \textbf{9.877} & \large \textbf{-1.859} & \large \textbf{1.420} & \large \textbf{-2.366} & \large \textbf{1.267} & \large \textbf{3.010} & \large \textbf{1.452} \\
    \large C2-SC   & \large 12.183 & \large 15.627 & \large -3.457 & \large 1.522 & \large -11.059 & \large 0.615 & \large 1.616 & \large 0.780 \\
    \large C2-USCC & \large 14.060 & \large 18.335 & \large -3.805 & \large 1.696 & \large -6.448 & \large 0.630 & \large 1.590 & \large 0.767 \\
    \large D2-SC   & \large 13.464 & \large 13.024 & \large -2.887 & \large 1.589 & \large -4.076 & \large 0.775 & \large 2.143 & \large 1.034 \\
    \large D2-USCC & \large \textbf{16.363} & \large 12.416 & \large -2.577 & \large 1.573 & \large -3.157 & \large 0.994 & \large 2.732 & \large 1.318 \\
    \hline
    \bottomrule
    \end{tabular}%
}}
\caption{Performance Metrics Table For Type 2 Models}
\label{tab:(12,1)_type_2_summary_stat}
\end{table}
\vspace{-0.6cm}

From Table~\ref{tab:(12,1)_type_2_summary_stat} we observe that, \textbf{B2(\ref{def:B2}) models} exhibit the best performance in terms of mean leverage, mean daily drawdown, Sharpe ratio and Expected Shortfall, followed by \textbf{D2(\ref{def:D2}) models}. Here the performance of \textbf{C2(\ref{def:C2}) models} is the worst. They have lower Sharpe value and ES value compared to other models. In terms of leverage, here also \textbf{A2(\ref{def:A2}) models} operate at a much higher leverage. Performance of models using \textbf{USCC} estimate are much better compared to the models using \textbf{SC} estimate of covariance matrix.

\subsubsection{Type-3 Models} \label{subsubsec:(12,1)-type-3-models}

\hspace{20pt}Now, we take a look at Type-3 models. We have models of 3 categories (A, B \& C) under Type-3 models. We will take a look at their performances.

\begin{table}[H]
\resizebox{\textwidth}{!}{%
{
    \setlength{\extrarowheight}{2pt}%
    \renewcommand{\arraystretch}{1.2}%
    \begin{tabular}{|l|ccc|cc|ccc|}
    \toprule
    \hline
    \multirow{2}{*}{\shortstack[c]{\textbf{Model}}} & 
    \multirow{2}{*}{\shortstack[c]{\textbf{Mean Monthly}\\\textbf{Return(\%)}}} & 
    \multirow{2}{*}{\shortstack[c]{\textbf{Monthly Return}\\\textbf{Std(\%)}}} & 
    \multirow{2}{*}{\shortstack[c]{\textbf{Monthly}\\\textbf{ES}}} & 
    \multirow{2}{*}{\shortstack[c]{\textbf{Mean}\\\textbf{Leverage}}} & 
    \multirow{2}{*}{\shortstack[c]{\textbf{Mean Daily}\\\textbf{Drawdown(\%)}}} & 
    \multicolumn{3}{|c|}{\textbf{Sharpe}} \\
    \cline{7-9}
    & & & & & & 
    \shortstack{\textbf{Daily}} & 
    \shortstack{\textbf{Weekly}} & 
    \shortstack{\textbf{Monthly}} \\
    \hline
    \large A3-SC   & \large 13.730 & \large 14.558 & \large -3.063 & \large 1.999 & \large -4.902 & \large 0.877 & \large 1.955 & \large 0.943 \\
    \large A3-USCC & \large \textbf{14.324} & \large 14.439 & \large -3.238 & \large 1.999 & \large -4.668 & \large 0.919 & \large 2.056 & \large 0.992 \\
    \large B3-SC   & \large 11.416 & \large \textbf{10.807} & \large \textbf{-2.344} & \large \textbf{1.510} & \large \textbf{-3.367} & \large 0.901 & \large 2.190 & \large 1.056 \\
    \large B3-USCC & \large 12.659 & \large \textbf{10.972} & \large \textbf{-2.303} & \large \textbf{1.649} & \large -3.520 & \large \textbf{1.002} & \large \textbf{2.392} & \large \textbf{1.154} \\
    \large C3-SC   & \large 13.357 & \large 11.586 & \large -2.475 & \large 1.816 & \large \textbf{-3.351} & \large 0.969 & \large 2.390 & \large 1.153 \\
    \large C3-USCC & \large \textbf{13.782} & \large 11.640 & \large -2.543 & \large 1.887 & \large -3.447 & \large \textbf{0.996} & \large \textbf{2.454} & \large \textbf{1.184} \\
    \hline
    \bottomrule
    \end{tabular}%
}}
\caption{Performance Metrics Table For Type 3 Models}
\label{tab:(12,1)_type_3_summary_stat}
\end{table}
\vspace{-0.6cm}

From Table~\ref{tab:(12,1)_type_3_summary_stat} we remark that, \textbf{B3(\ref{def:B3})} \& \textbf{C3(\ref{def:C3})} models dominate here, showing maximum Sharpe ratio and they also provide best drawdown among other Type-3 portfolio models. \textbf{B3} models show best performance in terms of ES value and mean leverage. \textbf{A3(\ref{def:A3}) models} offer the highest mean return but also exhibit greater return volatility, which lowers their daily and monthly Sharpe ratio to below 1. \textbf{C3 models} have the highest Sharpe ratio among type-3 models. However, considering Expected Shortfall (ES), \textbf{B3 and C3 models} are more or less comparable. \textbf{A3 models} consistently operate at high leverage, frequently reaching the threshold of 2. Here also, we see that models using \textbf{USCC} estimate perform slightly better than models using \textbf{SC} estimate.

\subsection{Training \& Testing Period (6,1)} \label{subsec:training-testing-period-(6,1)}

\hspace{20pt}Now, we will move on to the case where training period is 6 months and testing period is of 1 month. First we take a look at the Top-20 most selected stocks,

\begin{table}[H]
\centering
\caption{(6,1) Period Stock Selection Summary}
\label{tab:(6,1)-stock_selection}
\resizebox{\textwidth}{!}{%
  \setlength{\extrarowheight}{1pt}%
  \renewcommand{\arraystretch}{1.2}%
  \begin{tabular}{|c|c|c|}
    \toprule
    \hline
    \rowcolor{teal!70}
    \color{white}\textbf{Stock Name}
      & \color{white}\textbf{Sub-Industry}
      & \color{white}\textbf{Months Selected} \\
    \hline
    ExxonMobil
      & Integrated Oil \& Gas
      & 89 \\ \hline
    NVR, Inc.
      & Homebuilding
      & 65 \\ \hline
    Johnson \& Johnson
      & Pharmaceuticals
      & 63 \\ \hline
    McDonald's
      & Restaurants
      & 61 \\ \hline
    Waste Management
      & Environmental \& Facilities Services
      & 58 \\ \hline
    Southern Company
      & Electric Utilities
      & 53 \\ \hline
    U.S. Bancorp
      & Diversified Banks
      & 51 \\ \hline
    Walmart
      & Consumer Staples Merchandise Retail
      & 50 \\ \hline
    UnitedHealth Group
      & Managed Health Care
      & 44 \\ \hline
    Federal Realty Investment Trust
      & Retail REITs
      & 44 \\ \hline
    Home Depot (The)
      & Home Improvement Retail
      & 42 \\ \hline
    Coca-Cola Company (The)
      & Soft Drinks \& Non-alcoholic Beverages
      & 41 \\ \hline
    W. R. Berkley Corporation
      & Property \& Casualty Insurance
      & 40 \\ \hline
    Ecolab
      & Specialty Chemicals
      & 33 \\ \hline
    IBM
      & IT Consulting \& Other Services
      & 33 \\ \hline
    Lockheed Martin
      & Aerospace \& Defense
      & 33 \\ \hline
    Becton Dickinson
      & Health Care Equipment
      & 31 \\ \hline
    Public Storage
      & Self-Storage REITs
      & 30 \\ \hline
    3M
      & Industrial Conglomerates
      & 28 \\ \hline
    Arthur J. Gallagher \& Co.
      & Insurance Brokers
      & 28 \\
    \hline
    \bottomrule
  \end{tabular}%
}
\end{table}

Like before, here also the selection favors stable, well-established firms across industries, emphasizing resilience over high-growth potential. Strong defensive positioning, but limited tech and small-cap exposure. Now again we look at different types of models.

\subsubsection{Type-1 Models} \label{subsubsec:(6,1)-type-1-models}

\hspace{20pt}First, we take a look at Type-1 models. We have models of 4 categories (A, B, C \& D) under Type-1 models. We will take a look at their performances.

Just like the (12, 1) case of Type-1 models, here also we observe (from Table~\ref{tab:(6,1)_type_1_summary_stat}) that, \textbf{C1(\ref{def:C1}) models} perform the best in terms of mean monthly return, Monthly Expected Shortfall, Mean Daily Drawdown and Sharpe ratio, followed by \textbf{B1(\ref{def:B1}) \& D1(\ref{def:D1})} models. Here also \textbf{A1(\ref{def:A1}) models} operate at very high leverage values. For Type-1 models here, using \textbf{USCC} estimate for the covariance matrix yields slightly better results compared to \textbf{SC} estimate.
\vspace{0.2cm}

\begin{table}[H]
\resizebox{\textwidth}{!}{%
{
    \setlength{\extrarowheight}{2pt}%
    \renewcommand{\arraystretch}{1.2}%
    \begin{tabular}{|l|ccc|cc|ccc|}
    \toprule
    \hline
    \multirow{2}{*}{\shortstack[c]{\textbf{Model}}} & 
    \multirow{2}{*}{\shortstack[c]{\textbf{Mean Monthly}\\\textbf{Return(\%)}}} & 
    \multirow{2}{*}{\shortstack[c]{\textbf{Monthly Return}\\\textbf{Std(\%)}}} & 
    \multirow{2}{*}{\shortstack[c]{\textbf{Monthly}\\\textbf{ES}}} & 
    \multirow{2}{*}{\shortstack[c]{\textbf{Mean}\\\textbf{Leverage}}} & 
    \multirow{2}{*}{\shortstack[c]{\textbf{Mean Daily}\\\textbf{Drawdown(\%)}}} & 
    \multicolumn{3}{|c|}{\textbf{Sharpe}} \\
    \cline{7-9}
    & & & & & & 
    \shortstack{\textbf{Daily}} & 
    \shortstack{\textbf{Weekly}} & 
    \shortstack{\textbf{Monthly}} \\
    \hline
    \large A1-SC   & \large 12.774 & \large 11.762 & \large -2.739 & \large 1.750 & \large -3.716 & \large 1.028 & \large 2.251 & \large 1.086 \\
    \large A1-USCC & \large 13.199 & \large 11.517 & \large -2.679 & \large 1.787 & \large -3.579 & \large 1.075 & \large 2.376 & \large 1.146 \\
    \large B1-SC   & \large 11.489 & \large \textbf{10.459} & \large -2.561 & \large \textbf{1.574} & \large -3.305 & \large 0.921 & \large 2.277 & \large 1.098 \\
    \large B1-USCC & \large 12.470 & \large \textbf{10.305} & \large \textbf{-2.165} & \large 1.658 & \large \textbf{-3.276} & \large 1.020 & \large 2.508 & \large 1.210 \\
    \large C1-SC   & \large \textbf{16.005} & \large 10.702 & \large -2.167 & \large \textbf{1.615} & \large \textbf{-2.901} & \large \textbf{1.179} & \large \textbf{3.100} & \large \textbf{1.496} \\
    \large C1-USCC & \large \textbf{15.871} & \large 10.977 & \large \textbf{-2.109} & \large 1.817 & \large -3.356 & \large \textbf{1.161} & \large \textbf{2.997} & \large \textbf{1.446} \\
    \large D1-SC   & \large 12.620 & \large 10.625 & \large -2.416 & \large 1.638 & \large -3.363 & \large 1.008 & \large 2.462 & \large 1.188 \\
    \large D1-USCC & \large 12.735 & \large 10.719 & \large -2.544 & \large 1.785 & \large -3.462 & \large 1.014 & \large 2.463 & \large 1.188 \\
    \hline
    \bottomrule
    \end{tabular}%
}}
\caption{Performance Metrics Table For Type 1 Models}
\label{tab:(6,1)_type_1_summary_stat}
\end{table}

\subsubsection{Type-2 Models} \label{subsubsec:(6,1)-type-2-models}

\hspace{20pt}For Type-2 models also we look at their performance metrics table and from their we draw inference of the performance of different models.
\vspace{0.2cm}

\begin{table}[H]
\resizebox{\textwidth}{!}{%
{
    \setlength{\extrarowheight}{2pt}%
    \renewcommand{\arraystretch}{1.2}%
    \begin{tabular}{|l|ccc|cc|ccc|}
    \toprule
    \hline
    \multirow{2}{*}{\shortstack[c]{\large\textbf{Model}}} & 
    \multirow{2}{*}{\shortstack[c]{\large\textbf{Mean Monthly}\\\large\textbf{Return(\%)}}} & 
    \multirow{2}{*}{\shortstack[c]{\large\textbf{Monthly Return}\\\large\textbf{Std(\%)}}} & 
    \multirow{2}{*}{\shortstack[c]{\large\textbf{Monthly}\\\large\textbf{ES}}} & 
    \multirow{2}{*}{\shortstack[c]{\large\textbf{Mean}\\\large\textbf{Leverage}}} & 
    \multirow{2}{*}{\shortstack[c]{\large\textbf{Mean Daily}\\\large\textbf{Drawdown(\%)}}} & 
    \multicolumn{3}{|c|}{\large\textbf{Sharpe}} \\
    \cline{7-9}
    & & & & & & 
    \shortstack{\large\textbf{Daily}} & 
    \shortstack{\large\textbf{Weekly}} & 
    \shortstack{\large\textbf{Monthly}} \\
    \hline
    \large A2-SC   & \large 7.445 & \large 11.305 & \large -2.853 & \large 1.681 & \large -5.322 & \large 0.587 & \large 1.365 & \large 0.659 \\
    \large A2-USCC & \large 6.004 & \large 12.190 & \large -3.148 & \large 1.753 & \large -6.441 & \large 0.390 & \large 1.021 & \large 0.493 \\
    \large B2-SC   & \large \textbf{13.750} & \large \textbf{9.506} & \large \textbf{-2.092} & \large \textbf{1.269} & \large \textbf{-2.230} & \large \textbf{1.242} & \large \textbf{2.999} & \large \textbf{1.446} \\
    \large B2-USCC & \large \textbf{12.860} & \large \textbf{9.154} & \large \textbf{-2.061} & \large \textbf{1.388} & \large \textbf{-2.432} & \large \textbf{1.135} & \large \textbf{2.912} & \large \textbf{1.405} \\
    \large C2-SC   & \large 8.235 & \large 16.118 & \large -3.854 & \large 1.508 & \large -11.202 & \large 0.411 & \large 1.059 & \large 0.511 \\
    \large C2-USCC & \large 5.256 & \large 17.329 & \large -4.757 & \large 1.728 & \large -13.549 & \large 0.274 & \large 0.629 & \large 0.303 \\
    \large D2-SC   & \large 8.817 & \large 10.569 & \large -2.680 & \large 1.467 & \large -4.517 & \large 0.716 & \large 1.729 & \large 0.834 \\
    \large D2-USCC & \large 7.247 & \large 11.868 & \large -3.210 & \large 1.664 & \large -5.865 & \large 0.471 & \large 1.266 & \large 0.611 \\
    \hline
    \bottomrule
    \end{tabular}%
}}
\caption{Performance Metrics Table For Type 2 Models}
\label{tab:(6,1)_type_2_summary_stat}
\end{table}

From Table~\ref{tab:(6,1)_type_2_summary_stat}, we can infer that for this case, \textbf{B2(\ref{def:B2}) models} show best performance in terms of all performance measures, followed by \textbf{D2(\ref{def:D2}) \& A2(\ref{def:A2})} models. \textbf{C2(\ref{def:C2}) models} don't have good performance here. Also we see that, \textbf{A2} models operate at highest mean leverage. Here we see that contrary to previous cases, \textbf{SC} estimates show better performance compared to \textbf{USCC} estimates.

\subsubsection{Type-3 Models} \label{subsubsec:(6,1)-type-3-models}

\hspace{20pt}Finally, we look at the performance of Type-3 models. In the following we have the table of the performance metrics.

Based on Table~\ref{tab:(6,1)_type_3_summary_stat} of performance metrics, \textbf{B3(\ref{def:B3}) models} exhibit the best performance in terms of almost all of the metrics (except mean monthly return), followed by \textbf{C3(\ref{def:C3}) models}. We can see that here \textbf{A3(\ref{def:A3}) models} operate at highest amount of leverage available to the model. Here also, models using \textbf{USCC} estimate show slightly better compared to their equivalent models using \textbf{SC} estimates.

\begin{table}[H]
\resizebox{\textwidth}{!}{%
{
    \setlength{\extrarowheight}{2pt}
    \renewcommand{\arraystretch}{1.2}
    \begin{tabular}{|l|ccc|cc|ccc|}
    \toprule
    \hline
    \multirow{2}{*}{\shortstack[c]{\textbf{Model}}} & 
    \multirow{2}{*}{\shortstack[c]{\textbf{Mean Monthly}\\\textbf{Return(\%)}}} & 
    \multirow{2}{*}{\shortstack[c]{\textbf{Monthly Return}\\\textbf{Std(\%)}}} & 
    \multirow{2}{*}{\shortstack[c]{\textbf{Monthly}\\\textbf{ES}}} & 
    \multirow{2}{*}{\shortstack[c]{\textbf{Mean}\\\textbf{Leverage}}} & 
    \multirow{2}{*}{\shortstack[c]{\textbf{Mean Daily}\\\textbf{Drawdown(\%)}}} & 
    \multicolumn{3}{|c|}{\textbf{Sharpe}} \\
    \cline{7-9}
    & & & & & & 
    \shortstack{\textbf{Daily}} & 
    \shortstack{\textbf{Weekly}} & 
    \shortstack{\textbf{Monthly}} \\
    \hline
    \large A3-SC   & \large 11.004 & \large 14.277 & \large -3.180 & \large 2.000 & \large -5.607 & \large 0.691 & \large 1.598 & \large 0.771 \\
    \large A3-USCC & \large 11.518 & \large 13.897 & \large -2.993 & \large 2.000 & \large -5.329 & \large 0.735 & \large 1.718 & \large 0.829 \\
    \large B3-SC   & \large 11.895 & \large \textbf{10.026} & \large \textbf{-2.344} & \large \textbf{1.338} & \large \textbf{-2.990} & \large 0.981 & \large \textbf{2.459} & \large \textbf{1.186} \\
    \large B3-USCC & \large \textbf{12.028} & \large \textbf{9.689} & \large \textbf{-2.143} & \large \textbf{1.484} & \large \textbf{-2.963} & \large \textbf{0.999} & \large \textbf{2.573} & \large \textbf{1.241} \\
    \large C3-SC   & \large 11.259 & \large 12.805 & \large -2.911 & \large 1.960 & \large -4.742 & \large 0.769 & \large 1.823 & \large 0.879 \\
    \large C3-USCC & \large \textbf{14.147} & \large 12.471 & \large -2.752 & \large 1.975 & \large -4.039 & \large \textbf{0.991} & \large 2.352 & \large 1.134 \\
    \hline
    \bottomrule
    \end{tabular}%
}}
\caption{Performance Metrics Table For Type 3 Models}
\label{tab:(6,1)_type_3_summary_stat}
\end{table}

\section{Conclusion} \label{sec:conclusion}

\hspace{20pt}The findings of this study highlight the effectiveness of risk measures based on the equal-correlation portfolio strategy as a compelling alternative to traditional optimization methods. By constructing portfolios with a more balanced correlation structure, this approach enhances \textbf{risk diversification and return stability} across various market conditions.

\begin{itemize}
    \item Models incorporating correlation-based risk measures—\textbf{Categories B, C, and D} consistently \textbf{outperform conventional variance-based approaches (Category A)}.
    \item In Type-1 portfolios, \textbf{Category-C} models perform best, while in Type-2 and Type-3 portfolios, \textbf{Category-B} models lead.
    \item These results suggest that equalizing the correlation between individual asset returns and the portfolio return improves \textbf{risk-adjusted returns and downside protection}.
    \item The strategy is particularly suitable for institutional investors and systematic asset allocation frameworks.
\end{itemize}

While the stock selection method used is relatively simple, its emphasis on sectoral diversification, stability, and dividends contributes to the robustness of the proposed framework. Future research could explore more advanced selection techniques and their interaction with correlation-based portfolio construction. Lastly, we observed that in most cases, models using \textbf{USCC} estimates perform better than those using \textbf{SC} estimates of the covariance matrix.

\vspace{0.1cm}
\fancyline

\newpage

\sloppy
\bibliographystyle{plainurl}
\bibliography{references}

\end{document}